\documentclass[aps,twocolumn,nofootinbib]{revtex4}
\usepackage[latin1]{inputenc}
\usepackage{epsfig}
\usepackage{graphicx}
\newcommand{\beq}{\begin{equation}}
\newcommand{\eeq}{\end{equation}}

\begin{document}

\title{Reflections of Sadi Carnot}

\author{Mário J. de Oliveira}
\affiliation{Universidade de São Paulo,
Instituto de Física,
Rua do Matão, 1371, 05508-090
São Paulo, SP, Brasil}

\begin{abstract}

The Carnot theory is unique among the theories of heat developed
before the emergence of thermodynamics because it considers the
relationship between heat and work. The theory is contained in
Carnot's book published in 1824, which includes the basic ideas on
how thermal machines work, including the need for a temperature
difference. The fundamental principle of the theory is stated with
the help of a cyclic process invented by Carnot, involving two
isotherms and two adiabatics. The ratio between the mechanical work
produced during the cycle and the heat involved depends only on the
temperatures. We make a critical analysis of Carnot theory and show
how the fundamental principle was used by Clausius to define entropy
in terms of which he enunciated the second law of thermodynamics. 

\end{abstract}

\maketitle

\section{Introduction}

In 1824, Sadi Carnot published his {\it Réflexions sur la
Puissance Motrice du Feu et sur les Machines propres à Développer
cette Puissance} \cite{carnot1824},
a book containing his theory of heat. Motive power is Carnot's
term for the mechanical work performed by a heat engine. The
term fire refers to the heat involved in producing mechanical work.

The Carnot theory is unique among heat theories that appeared
before the emergence of thermodynamics because it addresses the
relationship between work and heat. To explain the production of
work by a heat engine, Carnot used an analogy with a falling body.
When a body falls from a greater height to a lower one, the work
done by the body is proportional to its mass and the difference
in height. When a certain amount of heat passes from a higher
temperature to a lower temperature, the work done is proportional
to amount of heat, and depends only on the two temperatures.

This explanation is the basis of the principle introduced by
Carnot, which is stated using a specific process through which
the body that performs work passes. This process is cyclic and
consists of two isothermal processes at different temperatures
and two processes that do not involve heat exchange. Using this
cyclic process, the principle is stated as follows:
\begin{quote}
The ratio of mechanical work to heat involved depends only on
the two temperatures and is independent of the nature of the
body performing the work.
\end{quote}

It should be emphasized that the heat involved is the heat
received by the body at a higher temperature, which is the
same heat released by the body at a lower temperature. Using
the analogy between the work produced by a heat engine and
the work done by a falling object, heat is analogous to the
mass of the body, which remains the same. In other words, heat
is a conserved quantity, which is the fundamental law of heat
theories prevalent before the emergence of thermodynamics,
called caloric theory \cite{fox1971}, implicit in the 
Carnot theory.

The Carnot theory was developed analytically by Clayperon
in the paper {\it Mémoire sur la puissance motrice de la
chaleur} published in 1834 \cite{clapeyron1834}. In this paper,
Clapeyron introduces the pressure-volume diagram to represent
the states of the system. He also obtains new results from the 
Carnot theory, such as the equation that evolves the
temperature and pressure of a liquid in equilibrium with its
vapor, which was later modified by Clausius and became known
as the Clausius-Clapeyron equation.

The Carnot principle was used by Clausius in formulating the
second law of thermodynamics, presented in his works on the
mechanical theory of heat, published from 1850 onwards \cite{clausius1850,clausius1854,clausius1865}. To this end,
Clausius separated the principle of Carnot into two parts.
He rejected the part concerning the conservation of heat as
incompatible with the conservation of energy and interpreted
the heat evolved as merely the heat introduced. Using the
Carnot principle, modified in this way, Clausius defined
entropy in terms of which he enunciated the second law of
thermodynamics.

Below, we present the Carnot theory in analytic form,
following the formulation of Clapeyron. 
After, we show how Clausius used the Carnot
principle to formulate the second law of thermodynamics.
We then, list the various editions of Carnot's book,
and we present a brief biography of Carnot, based
primarily on the biographical notes written by his brother
Hippolyte, contained in the second edition of
{\it Refléxions} published in 1878.

\section{Carnot theory}

\begin{figure}
\centering
\epsfig{file=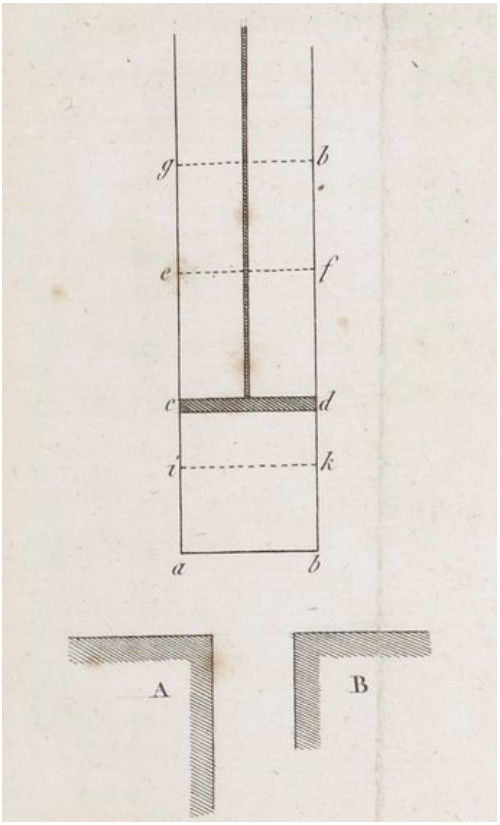,width=6cm}
\caption{Figure from the 1824 edition of {\it Réflexions} showing
the stages of the cycle invented by Carnot: 1) $cd$, initial position
of the piston; 2) $cd\to ef$, expansion in contact with body A;
3) $ef\to gh$, expansion in isolation, the temperature drops;
4) $gh\to cd$, compression in contact with B; 5) $cd\to ik$,
compression in isolation, the temperature rises;
6) $ik\to cd\to ef$, expansion in contact with body A.}
\label{maqui}
\end{figure}

We consider a body that undergoes a cyclic process consisting
of four stages, the Carnot cycle \cite[p. 32-34]{carnot1824}. 
The illustration of the cycle is given in figure \ref{maqui}.
An isothermal expansion at temperature $\theta_1$, an adiabatic
expansion, an isothermal contraction at temperature $\theta_2$,
and an adiabatic contraction. During the isothermal expansion,
the body is in contact with a reservoir A from which it receives
a quantity of heat $q$. In the adiabatic expansion, the body is
isolated. In the isothermal contraction, the body is in contact
with a reservoir B to which it gives off the heat $q$ received
from A. The work $w$ done by the body is given by
\beq
\frac{w}{q} = F(\theta_1,\theta_2),
\label{5}
\eeq
where $F(\theta_1,\theta_2)$ is a function that depends only on
the two temperatures and is therefore the same for any body.
In the language used by Carnot, it is independent of the nature
of the body. It is important to note that, in accordance with
the Carnot theory, the amount of heat
received by the body at temperature $\theta_1$ is the same
amount of heat given off by the body at temperature $\theta_2$,
both denoted by $q$, which is the result of the conservation
of heat. Equation (\ref{5}) analytically translates the 
Carnot principle \cite[p. 38]{carnot1824}.

If we consider a second Carnot cycle at temperatures $\theta_2$
and $\theta_3$ involving the same heat $q$, the work will be $w'=qF(\theta_2,\theta_3)$. For a third Carnot cycle at
temperatures $\theta_1$ and $\theta_3$ and with the same heat
$q$, the work will be $w''=qF(\theta_1,\theta_3)$.
Since $w''=w+w'$ then
\beq
F(\theta_1,\theta_3) = F(\theta_1,\theta_2) + F(\theta_2,\theta_3),
\eeq
from which we conclude that $F(\theta_1,\theta_2)$ has the form
\beq
F(\theta_1,\theta_2) = f(\theta_1) - f(\theta_2),
\eeq
where $f(\theta)$ depends only on $\theta$, and the principle
of Carnot takes the form
\beq
\frac{w}{q} = f(\theta_1) - f(\theta_2).
\eeq

If we consider a Carnot cycle formed by two very close isotherms
corresponding to the temperatures $\theta_1=\theta + d\theta$
and $\theta_2=\theta$, then the work $dw$ is given by
\beq
dw = q f' d\theta,
\label{2}
\eeq
where $f'=df/d\theta$.

Next, we consider any cyclic process and partition it into several
elementary cycles, each with a temperature difference equal to
$d\theta$. Each of these elementary cycles is approximated by a
Carnot cycle whose associated heat depends on the temperature.
The total work $w$ is the sum of the works of these elementary
cycles and therefore
\beq
w = \int q f' d\theta.
\label{27}
\eeq
The cyclic process is represented in the diagram $(v,p)$ as a
closed curve and the work $w$ is the area inside the curve,
\beq
w = \int_{\cal A} dp dv,
\eeq 
where ${\cal A}$ is the region bounded by the closed curve.
Analogously, the same cyclic process can be represented in
the diagram $(\theta,q)$ by a closed curve such that the
integral on the right-hand side of (\ref{27}) is given by
\beq
\int_{\cal B} f' dq d\theta,
\eeq
where ${\cal B}$ is the region bounded by the closed curve
in the plane $(\theta,q)$.

Using the results above we write
\beq
\int_{\cal A} dp dv = \int_{\cal B} f' dq d\theta,
\label{36}
\eeq
so that $f'$ is understood as the Jacobian of the
transformation $(v,p)\to(\theta,q)$ and therefore
\beq
\frac{\partial q}{\partial v}\frac{\partial\theta}{\partial p}
- \frac{\partial \theta}{\partial v}\frac{\partial q}{\partial p} = h,
\label{30}
\eeq
where $h=1/f'$ is a function of $\theta$.

Equation (\ref{30}) can be understood as the fundamental equation
of Carnot theory. From it we can determine $q(v,p)$ once we know
$\theta(v,p)$. For example, for a gas,
\beq
pv=R(\theta+a).
\label{13}
\eeq
Carnot used the value $a=267$ and Clapeyron the value $a=273$. Using this equation, we obtain
\beq
v\frac{\partial q}{\partial v} 
- p \frac{\partial q}{\partial p} = R h,
\label{31}
\eeq
which is the equation derived by Clapeyron in his 1834 paper.

The universal function $h$, which Clapeyron called $C$, plays a
fundamental role in Carnot theory. In his 1834 paper, Clapeyron
states that ``it is the common place of all the phenomena produced
by heat in solid, liquid, or gaseous bodies. It would be desirable
for highly precise experiments, such as research on the propagation
of sound in gases carried out at different temperatures, to make
this function known with all the desired precision. It would serve
to determine several other important elements of the theory of heat.''

Considering that $h$ is a function of $\theta$ and therefore that
it is a function of the product $pv$, equation (\ref{31}) can be
integrated. The result is
\beq
q = R h \ln \frac{v}{v_0},
\label{22}
\eeq
where $v_0$ depends on $\theta$. Carnot refers to this result
in the following terms. The heat absorbed along an isotherm increases
in arithmetic progression when the volume increases in geometric
progression \cite[p. 52]{carnot1824}.

Then we write (\ref{22}) in the form
\beq
q = g\,p_0v_0 \ln\frac{v}{v_0},
\eeq
where $g(\theta)=h(\theta)/(\theta+a)$ only depends on $\theta$.
This result tells us that different gases consume the same heat
when traversing an isotherm if they initially have the same
temperature, the same pressure and the same volume
\cite[p. 41-42]{carnot1824}.
We note that for this to be possible, the masses of the gases
must be different.

Taking the derivative of (\ref{22}) at constant $v$, we find
the heat capacity at constant $v$, defined by $c_v=(dq/d\theta)_v$,
\beq
c_v = R h' \ln \frac{v}{v_0},
\eeq
where $h'=dh/d\theta$, and deriving the same expression at
constant $p$, we find the heat capacity at constant $p$,
defined by $c_p=(dq/d\theta)_p$,
\beq
c_p = R h' \ln \frac{v}{v_0} + \frac{R h}{\theta+a}. 
\eeq
When the volume increases geometrically, the heat capacities
increase arithmetically \cite[p. 58]{carnot1824}. This result means
that, for a given mass of gas, the heat capacities depend on the volume,
\beq
c_p-c_v = \frac{R h}{\theta+a},
\eeq
is independent of volume, that is, independent of density
\cite[p. 58]{carnot1824}.

When a gas expands due to a change $\Delta \theta$ in temperature,
the change in volume will be proportional to $\Delta\theta$ for
small values of that change. If the expansion is done at constant
pressure, the original volume $v$ will become
$v'=v+v\alpha\Delta\theta$. If the expansion is done without heat
intervention, then the new volume will be
$v'=v-v\beta\Delta \theta$. For a gas, the equation of state
(\ref{13}) gives $\alpha=1/(\theta+a)$. Therefore, at
$0^\circ$, $\alpha=1/a$, or $\alpha=1/267$, which is the value
assumed by Carnot \cite[p. 44]{carnot1824}. As for $\beta$, Carnot
assumes the value $\beta=1/116$ obtained experimentally by the sudden
compression of a gas \cite[p. 43]{carnot1824}. Carnot claims that this
result is due to Poisson and that this value agrees well with
the results of Clément and Desormes \cite[p. 43]{carnot1824}.

Carnot argues \cite[p. 42-45]{carnot1824} that $\alpha$ and $\beta$
are related to the ratio of heat capacities in the following way
\beq
\frac{c_p}{c_v} = 1-\frac{\alpha}{\beta}.
\label{35}
\eeq
Using the numerical results, Carnot arrived at the
result \cite[p. 45, 60]{carnot1824}
\beq
\frac{c_p}{c_v} = \frac{267+116}{267} = 1.44,
\eeq
and therefore close to the value 1.3748 obtained experimentally by Gay-Lussac and Welter, cited by Carnot \cite[p. 59]{carnot1824}.

The identity (\ref{35}) is proved as follows. We start with
the identity
\beq
\left(\frac{dq}{d\theta}\right)_p = 
\left(\frac{dq}{d\theta}\right)_v + 
\left(\frac{dq}{dv}\right)_\theta \left(\frac{dv}{d\theta}\right)_p,
\eeq
from which we obtain
\beq
c_p - c_v = c_v
\frac{(dq/dv)_\theta}{(dq/d\theta)_v}\left(\frac{dv}{d\theta}\right)_p,
\eeq
which can be written as
\beq
\frac{c_p}{c_v} - 1 = - \frac{(dv/d\theta)_p}{(dv/d\theta)_q}.
\eeq
But $\alpha=(1/v)(dv/d\theta)_p$ and $\beta=-(1/v)(dv/\theta)_q$
and we arrive at the result (\ref{35}). It is worth noting that
the relation (\ref{35}) is also valid within the domain of
thermodynamics. Although Carnot's theory and thermodynamics
lead us to different results, generally speaking, some results
may be the same as it happens with the relation (\ref{35}) and
also with the relation
\beq
\frac{c_p}{c_v} = \frac{\kappa_T}{\kappa_q},
\eeq
which involves compressibilities
$\kappa_T=-(1/v)(dv/dp)_\theta$ and
$\kappa_q=-(1/v)(dv/dp)_q$, known long before the
emergence of thermodynamics.

Let us consider a liquid in equilibrium with its vapor and a
Carnot cycle. Since phases coexist throughout a process at
constant temperature, the pressure is also constant. Therefore,
the work along an isotherm is equal to $p(v_g-v_l)$ where
$v_g$ and $v_l$ are the volumes of the vapor and liquid.
Therefore, considering two similar temperatures, then
$dw=dp(v_g-v_l)$ which, substituted into (\ref{2}),
leads us to the result
\beq
\frac{dp}{d\theta} = \frac{q f'}{(v_g-v_l)},
\eeq
which is the equation derived by Clayperon \cite{clapeyron1834}.
This equation was later modified by Clausius within the scope
of thermodynamics and became known as the Clausius-Clapeyron
equation.

It is interesting to note that equation (\ref{36}) can be
written in the form
\beq
\oint (f dq-pdv) = 0,
\eeq
where the left-hand side is an integral along a closed path
in the diagram $(v,q)$. Therefore, the integral
\beq
\int_{\cal C} (fdq - pdv),
\eeq
between two points of the diagram $(v,q)$ is independent of
the path ${\cal C}$, which means that we can define a
quantity $\phi$ such that
\beq
d\phi = fdq - pdv,
\eeq
or in other terms, $fdq-pdv$ is an exact differential.
Although $pdv$ is infinitesimal work, $fdq$ is not
infinitesimal heat since infinitesimal heat is $dq$.

\section{Entropy and the second law of thermodynamics}

The Carnot principle was used by Clausius in formulating the
second law of thermodynamics, presented in his works on the
mechanical theory of heat, published from 1850 onwards
\cite{clausius1850,clausius1854,clausius1865}. To understand
how Clausius did this, we state the Carnot principle in two parts:
\begin{quote}
1. A body performs work by receiving a quantity of heat from the outside and releasing the same quantity of heat to the outside. \\
2. The ratio between the mechanical work performed by the body and the heat received depends only on the two temperatures and is independent of the nature of the body performing the work. \\
\end{quote}
The first part is incompatible with the conservation of energy
established by Mayer and Joule during the 1840s. Clausius rejected
part 1 and retained part 2. In other words, Clausius interpreted
the heat evolved, which appears in Carnot's original statement, only
as the heat received. Thus, Clausius could assert that the heat given
off is different from the heat received and is such that their
difference is equal to the work done, in agreement with the
conservation of energy.

Using the Carnot cyclic process, the conservation of energy is
expressed as follows. A certain amount of heat $Q_1$ is received
by the body undergoing the cycle from a body A at temperature
$\theta_1$. Part of it is transformed into work $W$, and the
remaining part $Q_2'$ is transferred to a body B at temperature
$\theta_2$. The conservation of energy is therefore written as
\beq
W=Q_1-Q_2'.
\label{6}
\eeq 
Having rejected the first part of the Carnot principle and kept
the second one, this amounts to replacing $q$ in the expression
(\ref{5}) by the introduced heat $Q_1$, which takes the form
\beq
\frac{W}{Q_1}=F(\theta_1,\theta_2).
\eeq
Substituting (\ref{6}) into this expression, it becomes
\beq
\frac{Q_2'}{Q_1}=f(\theta_1,\theta_2),
\label{7}
\eeq
where $f=1-F$ and depends only on $\theta_1$ and $\theta_2$.

If we consider a second Carnot cycle at temperatures
$\theta_2$ and $\theta_3$, and denote by $Q_3'$ the heat
given off, then $Q_3'/Q_2'=f(\theta_2,\theta_3)$. If we consider
a third cycle composed of the first two, then
$Q_3'/Q_1'=f(\theta_1,\theta_3)$. Since
$Q_3'/Q_1'=(Q_3'/Q_2')(Q_2'/Q_1)$ then

\beq
f(\theta_1,\theta_3)= f(\theta_1,\theta_2)f(\theta_2,\theta_3).
\eeq
From this expression we see that the function $f$ has the
form $f(\theta_1,\theta_2)=\phi(\theta_2)/\phi(\theta_1)$,
where $\phi(\theta)$ depends only on $\theta$. Defining
$T=\phi(\theta)$, we can write
\beq
\frac{Q_2'}{Q_1}=\frac{T_2}{T_1}.
\eeq
Putting $Q_2=-Q_2'$, then
\beq
\frac{Q_1}{T_1}+\frac{Q_2}{T_2} = 0.
\eeq
Considering any cyclic process approximated by several
Carnot cycles, then this relation is written
\beq
\sum_i \frac{Q_i}{T_i} = 0,
\eeq
or, in the limit where the heats exchanged are infinitesimal,
\beq
\oint\frac{dQ}{T} = 0.
\label{12}
\eeq
This closed integral means that $dQ/T=dS$ is an exact
differential and therefore there exists a state function
$S$ that Clausius calls entropy. If we consider any process,
the entropy change is given by
\beq
\Delta S = \int \frac{dQ}{T},
\label{12a}
\eeq
and is path independent.

In the expressions (\ref{12}) and (\ref{12a}) the temperature
$T$ is the temperature of the body, which is equal to the
temperature of the environment with which the body exchanges heat.
That is, the process occurs in such a way that at any instant
the body has the same temperature as the environment. However,
in general, the temperatures are different. In his book, Carnot
comments that it is necessary for the temperature differences to
be as small as possible to extract the maximum possible work.
Let us assume that the temperature of the environment is $T'$.
For a quantity of heat to enter the body, it is necessary that
$T'\geq T$, which leads us to the result
\beq
dS = \frac{dQ}{T}\geq \frac{dQ}{T'},
\eeq
which can be understood as the expression of the second law of
thermodynamics in differential form. Integrating over a
process we obtain
\beq
\Delta S \geq \int\frac{dQ}{T'},
\eeq
which is the expression of the second law of thermodynamics
in the formulation given by Clausius in integrated form.

\section{Book}

\begin{figure}
\centering
\epsfig{file=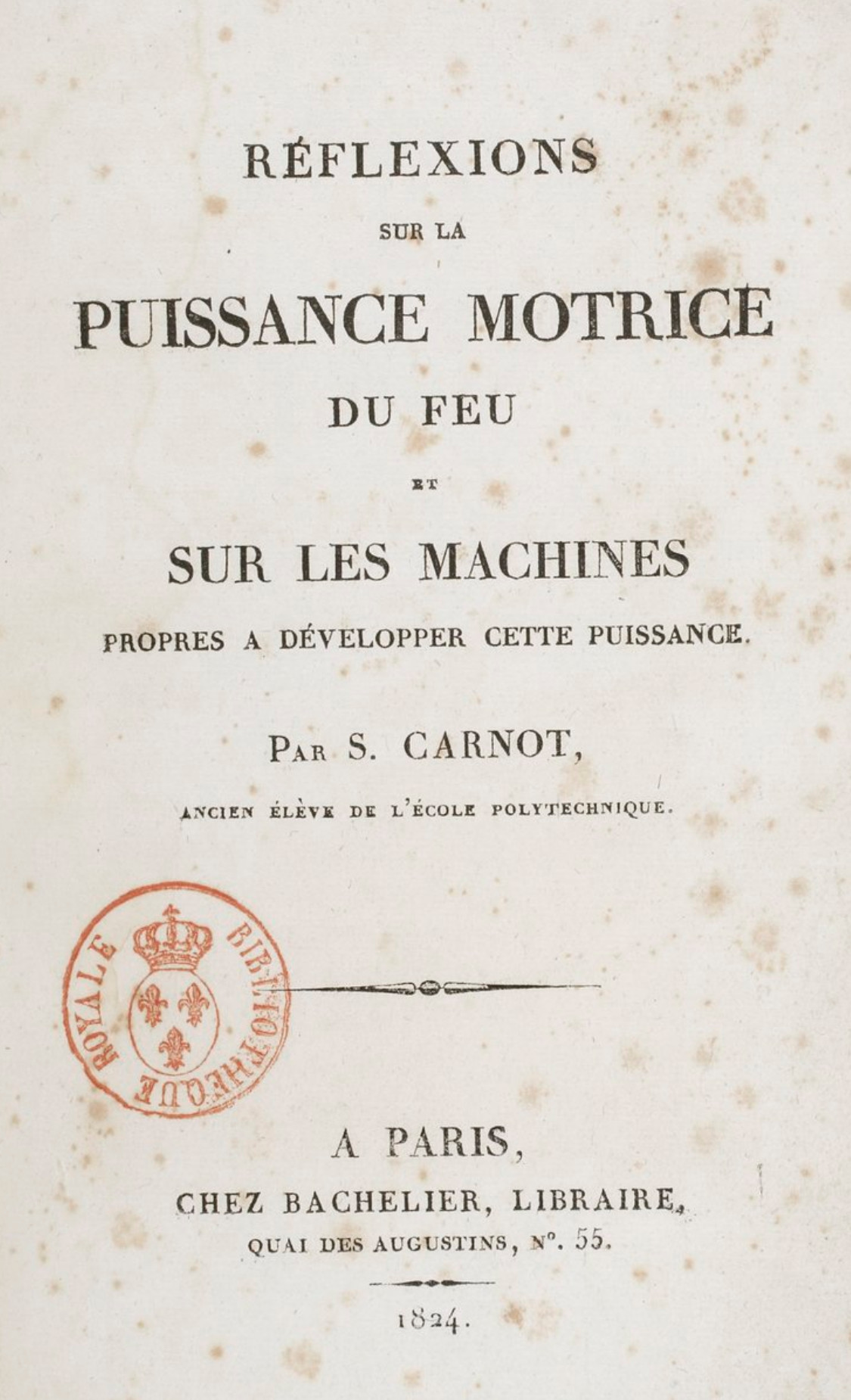,width=8.5cm}
\caption{Title page of the 1824 edition of Sadi Carnot's
{\it Réflexions}.}
\label{rosto1824}
\end{figure}

The title page of Carnot's {\it Réflexions} can be seen in figure
\ref{rosto1824}. The term {\it puissance motrice}, meaning work,
was used exceptionally by Carnot since the more common terms were,
for example, {\it force motrice} and {\it quantité d'action}.
The term {\it travail} was introduced by Coriolis in 1829.
Although Clapeyron wrote {\it puissance motrice} in the title of
his 1834 paper \cite{clapeyron1834}, the term he used within the
paper for work is {\it quantité d'action}.

The theory of heat presented in {\it Réflexions} is unique among
those that emerged during the period when the caloric theory
prevailed and therefore before the emergence of Clausius's
mechanical theory of heat. Despite the uniqueness and relevance
of Carnot's theory, his book received little attention in the two
decades following its publication. The significant exception is
Clapeyron, who based his 1834 paper on the Carnot theory
\cite{clapeyron1834} and explicitly refers to Carnot's book.

Clapeyron's paper was translated into English in 1837
\cite{clapeyron1837} and partially into German in 1843
\cite{clapeyron1843}. In the preface to the German translation,
the editor clarifies that the paper received little attention
but is being published because of its importance. In his 1845
book on the heat and elasticity of gases \cite{holtzmann1845},
Holtzmann cites Clapeyron's paper and states in the preface that
the paper is based on Carnot's work, but that he was unable to
obtain a copy of Carnot's book.

In 1848, Thomson published a paper \cite{kelvin1848} on th
absolute temperature scale based on the Carnot theory. He stated 
in this paper that he had not found Carnot's original book but
had reached its content through the paper by Clapeyron, whom
he had met in 1845 during his time in Regnault's laboratory.
However, in an 1849 paper \cite{kelvin1849} on the Carnot theory,
Kelvin already made direct reference to Carnot's book.

Clausius also refers to Carnot in his 1850 paper on the theory
of heat \cite{clausius1850}, saying that he could not find a
copy of the book and that he had become familiar with Carnot's
ideas through the work of Clapeyron and Thomson. In an 1863
paper \cite{clausius1863}, Clausius refers to Carnot's results
by explicitly citing his book.

\begin{figure}
\centering
\epsfig{file=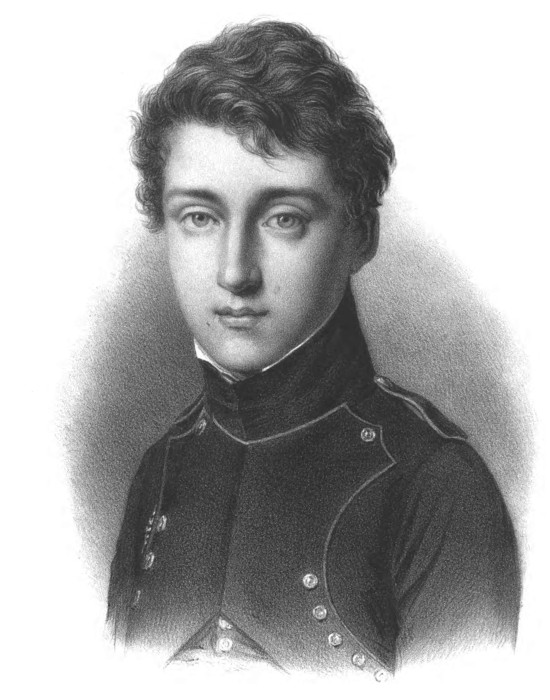,width=8cm}
\caption{Sadi Carnot, aged 17, in École Polytechnique uniform,
according to a portrait painted by Boilly in 1813, appearing
in the 1878 edition of {\it Réflexions}.}
\label{boilly}
\end{figure}

The Carnot treatise later appeared in the Annales Scientifiques
de l'École Normale Supérieure in 1872 \cite{carnot1872}. This
text was used in the 1878 publication \cite{carnot1878}, which also
contained three supplements. The first is a letter from his brother
Hippolyte Carnot, a senator of the Republic, addressed to the Academy
of Sciences and to the President of the Republic, François Sadi Carnot,
Carnot's nephew and son of Hyppolyte. The second is biographical notes
on Carnot written by Hippolyte, and the third is an extract from
Carnot's handwritten notes preceded by a facsimile of these notes
concerning the mechanical equivalent of heat. The book also contains
a reproduction of a portrait of Carnot at age 17 in the uniform of
the École Polytechnique, as seen in figure \ref{boilly}.

\begin{figure}
\centering
\epsfig{file=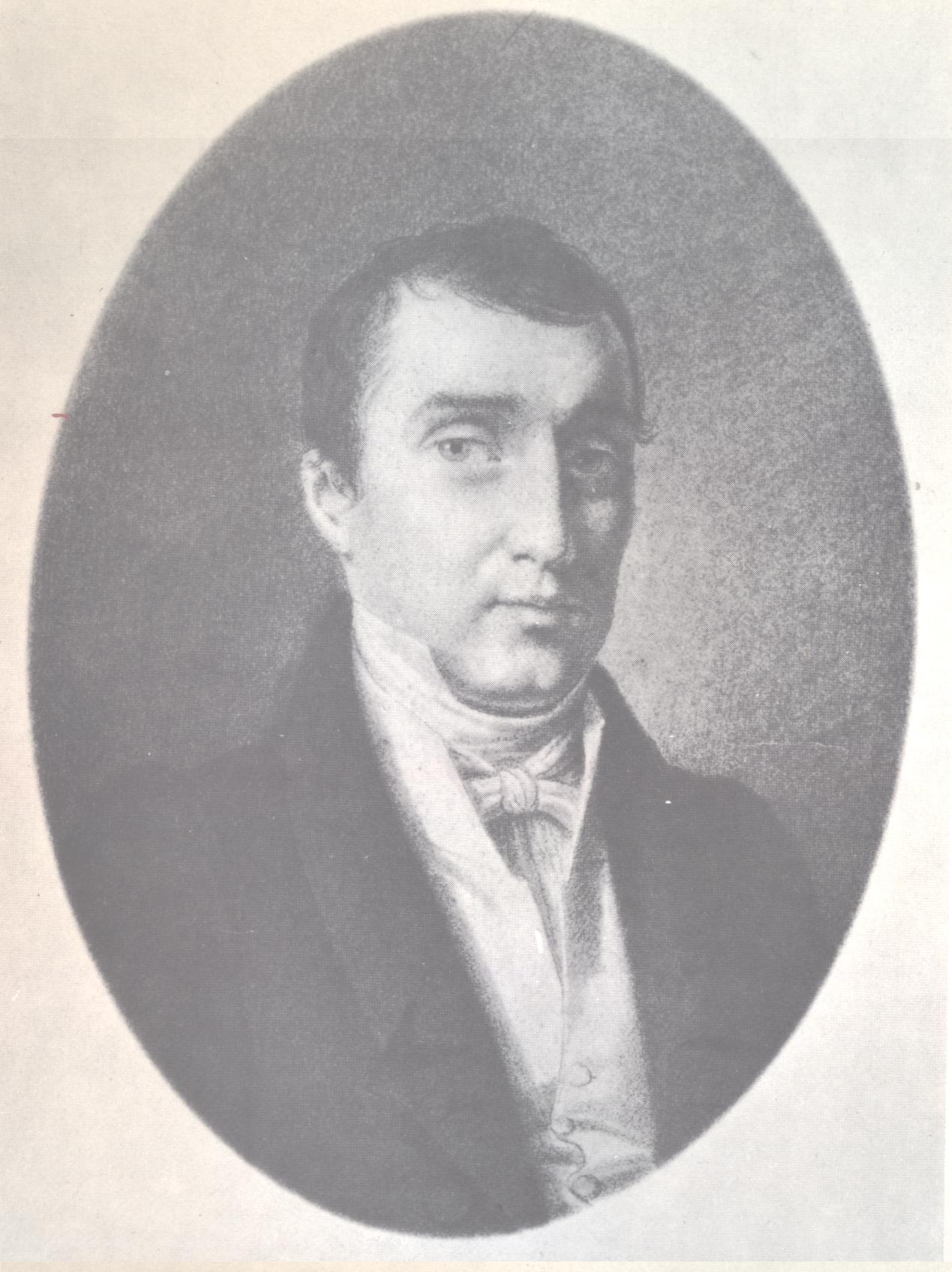,width=8cm}
\caption{Portrait of Sadi Carnot painted by Despois in
1830, contained in the 1978 critical edition.}
\label{despois}
\end{figure}

There is a 1903 edition \cite{carnot1903} that reproduces in facsimile
form the 1824 text and is supplemented by a reproduction of a sheet
of Carnot's handwritten notes concerning the mechanical equivalent of
heat. The 1953 edition \cite{carnot1953} is also a facsimile
reproduction of the 1824 text and has as an appendix Hippolyte
Carnot's letter and Carnot's scientific manuscripts. In 1978,
a critical edition of the 1824 French text was published by Fox
\cite{carnot1978} containing documents and several of Carnot's
manuscripts. This edition also contains a reproduction of an 1830
portrait of Carnot by Despois, as seen in figure \ref{despois}.

The English translation of Carnot's book was published in 1890
\cite{carnot1890} and a second edition in 1897 \cite{carnot1897}.
This second edition also contains Kelvin's 1849 paper
\cite{kelvin1849} on the Carnot theory. There is also another
English translation published in 1899 \cite{magie1899}, which
also contains the English version of Clausius's 1850 paper
\cite{clausius1850} and Kelvin's 1851 paper \cite{kelvin1851}.
Another English edition followed in 1943 \cite{carnot1943}.
It is also worth mentioning the 1960 publication \cite{mendoza1960}
which, in addition to the English translation of Carnot's book,
also contains the English translations of Clapeyron's 1834 paper 
\cite{clapeyron1834} and Clausius's 1850 paper \cite{clausius1850}.
In 1986, the English translation of Fox's critical edition mentioned
above appeared \cite{carnot1986}.

The German translation of Carnot's book was carried out by Ostwald and
published in 1892 \cite{carnot1892} as issue 37 of the collection
{\it Ostwald's Klassiker der Exakten Wissenschaften}. The Russian
translation was published in 1923 \cite{carnot1923}. There is a
Spanish translation from 1927 \cite{carnot1927} and another from
1987 \cite{carnot1987}. In 1992, an Italian edition appeared
\cite{carnot1992}.

\section{Bibliographic notes}

Nicolas Léonard Sadi Carnot was born in Paris on June 1, 1796.
His father was Lazare Carnot, known as the Organizer of Victory
for his role in the French Revolution. He was one of the five
Directors who held executive power in France between 1795 and
1799. As a member of the Directory, he lived in
{\it petit Luxembourg}, where Sadi was born. This name was due
to Lazare's fondness for the thirteenth-century Persian poet.
After the coup d'état of 1797, Lazare went
into exile, while Sadi and his mother, Sophie Dupont, lived
in Saint-Omer, near Calais.

With Napoleon's rise to power, Lazare returned to France and
briefly became a minister in 1800. In addition to his political
and military activities, Lazare also devoted himself to science.
Between 1801, when Sadi's brother, Hippolyte, was born, and 1813,
the year of Sophie's death, Lazare devoted himself intensely to
mathematics and mechanical science. In 1803, he published the
second edition of his book on the theory of machines
\cite{lazare1803}. During this period, Lazare also took charge
of Sadi's education until he turned sixteen \cite{carnot1978}.
After eight months of preparation, attending Bourdon's course at
the Lycée Charlemagne, Sadi entered the École Polytechnique in
November 1812, at only sixteen years of age.

Life for the school's boarders was very restrictive, with strict
military discipline. They rose at five o'clock, attended classes
for nine hours a day, except Sundays, and were monitored for
everything they did. It's hard to imagine anyone happy under
this regime, particularly Sadi. From December 1813 onward, the
polytechnics prepared militarily for the defense of Paris in
March 1814. Sadi bravely participated in the military action at
Vincennes, although Paris ultimately fell to Coalition troops.

In October 1814, Sadi completed his studies at the École
Polytechnique and in January 1815 entered the School of
Artillery and Engineering Application in Metz. He remained
there until being appointed second lieutenant in 1817.
For the next two years, he served in the military engineering
service. In 1819, he retired from this service upon being appointed
lieutenant of the General Staff. This allowed him to obtain
permanent leave and reside for over ten years in Paris. Freed
from the constraints of military life, he was able to begin the
study and research activities that continued until the end of
his life. However, it appears that he never held or sought an
academic position or a teaching position.

In 1821 Sadi interrupted his studies to visit his father,
who was in exile in Magdeburg. It is quite possible that
conversations with his father on this occasion drew Sadi's
attention to the problems addressed in the {\it Réflexions}.

According to his brother Hippolyte, Sadi attended the
Conservatoire des Arts et Métiers. In 1819, the Conservatoire
established three chairs, including one in applied chemistry,
entrusted to Nicolas Clément. He and Charles Desormes are known
for their 1819 experiments to determine the ratio between the
heat capacities of gases at constant pressure and constant
volume. These experiments are based on the adiabatic
expansion of gases.

Some notes from Clément's lectures indicate that he was
also developing a theory of heat engines at the same time
as Sadi. This suggests that Clément influenced Sadi. The
greatest influence may have been on the explanation of the
free expansion of a gas. According to Clément and Desormes,
sudden expansion is understood as a process that occurs
without heat exchange and not isothermally, so that the
temperature decreases.

In addition to his scientific interests, it is possible that
Sadi also had industrial interests, which is consistent with
his concern for the progress of French industry, as expressed
in the introduction of the {\it Réflexions}. Sadi was one of
the first members of the {\it Association Polytechnique},
founded shortly after the 1830 revolution for the purpose
of popular dissemination of scientific and technological
knowledge.

Sadi died on October 24, 1832, at the age of 36. In the
biographical notes of the 1878 edition of {\it Réflexions},
we find the following message written by Hippolyte. 
``Sadi Carnot died in the prime of his life, on the threshold
of a career he promised to achieve with brilliance, leaving a
memory of deep esteem and affection in the hearts of some
friends. His manuscripts attest to the activity of his mind,
the variety of his knowledge, his love for humanity, his
enlightened sentiments of justice and freedom. One can follow
the traces of all kinds of studies in them. But the only work
he completed is the one we publish for the second time. It will
be enough to ensure that his name is not forgotten.''


\end{document}